\documentstyle[preprint,aps,epsf]{revtex}

\newcommand{\be}{\begin{equation}}
\newcommand{\bfig}{\begin{figure}}
\newcommand{\ee}{\end{equation}}
\newcommand{\efig}{\end{figure}}
\newcommand{\bi}{\begin{itemize}}
\newcommand{\ei}{\end{itemize}}
\newcommand{\bear}{\begin{eqnarray}}
\newcommand{\eear}{\end{eqnarray}}
\newcommand{\ba}{\begin{array}}

\newcommand{\ea}{\end{array}}

\begin{document}

\title {A microsimulation of traders activity  in the stock market: \\
the role of heterogeneity, agents' interactions and trade frictions.}
\author{Giulia  Iori } 
\address{  Department of Accounting, Finance and Management, \\
University of Essex\\
 Wivenhoe Park, Colchester CO4 3SQ, UK  \\
{\small  iorig@essex.ac.uk}}
\date{\today}
\maketitle

\begin{abstract}

We propose  a model with heterogeneous interacting traders  
which can explain some of the stylized facts of
stock market returns. 
In the  model,  synchronization effects, which generate large
fluctuations in  returns,  can  arise  purely  from communication and
imitation among traders. The key element in the model is the
introduction of a trade friction  which, by  responding
 to price movements, creates a   feedback mechanism
 on  future trading and generates volatility clustering. The
model also reproduces the  empirically observed  positive
cross-correlation between volatility and trading volume.

\vskip 1cm
\end{abstract}
{\bf JEL classification:} G12, C63\\
{\bf Keywords:} Volatility clustering, fat tails, trading volume, herd behaviour.\\

\newpage

\section{Introduction}

As pointed out  by several authors, since Mandelbrot (1963),
 the probability distributions of returns 
 of many market indices and currencies, over different
 but relatively short time intervals, display fat tails.
It is estimated that the shape of a Gaussian, as predicted by the
 random-walk hypothesis, is  recovered only 
on longer time scales, i.e. of  one month or more.
The empirical distribution of returns 
shows  an asymptotic power law decay
 $P(r) \sim r^{-(1 + \mu)}$
 with an exponent  $2 < \mu < 4$
(Pagan (1996),  Guillaume et al. (1997), Gopikrishnan et al. (1999)).  

Moreover,  while stock market returns are
uncorrelated on lags larger than a single day, the autocorrelation
function  of the volatility is positive and slowly
decaying, indicating  long memory effects.
 This phenomenon is known in the literature as  volatility clustering
(Ding et al. (1993), de Lima \& Crato (1994), Ramsey (1997), Ramsey \&
Zhang (1997)).  
Empirical analysis  on market  indices and exchange rates
shows  that  the volatility autocorrelations  are 
power-laws over a large  range of time lags (from one day to one year),
in contrast with ARCH-GARCH models  (Engle (1995),
 Bollerslev (1986)) where they are
supposed to  decay exponentially.

Daily financial time series also provide empirical evidence 
(Tauchen \& Pitts (1983), Ronals et al. (1992), Pagan (1996)) of a
positive autocorrelation, with slowly decaying tails,  for
the trading volume, and  positive cross correlation
between  volatility of returns  and   trading volume.

There is also some evidence 
 that both the moments of the distribution of returns
 (Ghashghaie et al. (1996), Baviera et
 al. (1998))  and the volatility
 autocorrelations (Baviera et
 al. (1998), Pasquini \& Serva (1998))
  display what is termed in the theory  of turbulent systems
as  multiscaling (see section IV for a description of multiscaling).

 It is not settled yet
whether  these power law fluctuations
merely reflect the probability distributions of exogenous shocks
hitting the market or whether they are related 
to the inherent interaction among market players and the trading
process itself. 
 For example, it is widely believed that stock markets are vulnerable
 to  collective behaviour whereby       
 a large group of agents place the  same order simultaneously. This can 
 manifest  itself  in  large price 
fluctuations, and eventually  crashes.
Such collective behaviour could reflect the phenomenon known as herding
 which occurs when agents take
actions on the basis of imitating each other 
(Bannerjee (1992), Bannerjee (1993), Orlean (1995), Cont \& Bouchaud (1999)).
To model how the decisions  of agents are influenced by
their mutual interaction,
this paper assumes a simple communication structure based on ideas from 
statistical physics
in which  agents occupy the nodes of a lattice and are influenced by the
decisions of their nearest neighbours. 
Agents are assumed to be heterogenous in their exposure to a trading
friction which can restrict market participation.
Heterogeneity ensures that traders do not spontaneously make  the same decisions; 
indeed, this would   make superfluous the coordinating role of communication.
We pursue the claim that  the  interactions among traders 
and their  heterogeneous nature by themselves, 
 can help explain  the statistical features observed in  the  financial
data.
Since the crucial point is not 
 the  exact description of individual behaviour but the interrelation
between  and the statistical differences across agents, our model
will be based on
 noise trading expanded to allow for heterogeneity and communication effects.

A variety of  agent-based model
have been proposed in the last few years 
 to study financial markets ( Arthur (1997), Caldarelli
et al. (1997), Bak et al. (1997), Farmer (1998), LeBaron (1999),
 Lux and Marchesi (1999), Cont \& Bouchaud (1999),
 Focardi et al. (1999), Stauffer et al. (1999)). 
Cont and Bouchaud (1999) introduce a  model which is similar to the
one we use in this paper. One of the differences is that in their model 
imitation takes the form that agents act in clusters rather than
individually. Furthermore, a probability $p$ is associated with a link
existing between any two agents and this determines the
distribution of cluster sizes.
The trading activity of each  group is controlled  exogenously.
 For low activity and  at a critical value of $p$ 
the distribution of returns shows a power law behaviour
with an exponent $\mu$ which can vary from $1$ to $1.5$ (depending on the
dimensionality of the system). This is smaller  than the empirically observed
value $\mu \sim 3$ which our own model is capable of reproducing.
Various mechanisms have been proposed to  improve the
agreement of the Cont-Bouchaud model with real data. 
Zhang (1999)  suggests  a non 
linear price rule which   assumes that price changes are
proportional to the square root of the
difference between demand and supply. Zhang finds an exponent  $\mu = 1.9$ 
for  a two dimentional lattice, which is still below the
 empirical value. 

Both the above papers are concerned only with
explaining the power-law behaviour of the distribution of  returns and
not with volatility clustering. 
In a recent paper, Stauffer and Sornette (1999) consider a modified 
version of the Cont-Bouchaud model,  
where the connectivity of the system is  changed by slowly adjusting the
value of  $p$ at each  time step.
This mechanism produces volatility clustering and a better
agreement  with the empirical exponent. However, the adjustment of  $p$
is entirely exogenous and does not have an underlying economic explanation.

Our model suggests an endogenous mechanism
for changing the level of correlation among agents (analogous to changing the
connectivity in the above paper) and also endogenizes the exponent of
the pricing rule (which is held constant to $1/2$  in  Zhang).
It provides a better agreement with the stylized
facts of stock market dynamics, including cross-correlation between
trading volume and volatility which were not addressed by the other authors. 

The rest of the paper is organized as follows:  the next section
describes the model and the simulation strategy, section $3$ discusses
results and section $4$ concludes.

\section{The Model}

A modified version of the random field Ising model (RFIM)
(Sethna (1993), Galam(1997))
is employed  to describe the trading behaviour of  agents in a stock
market.
We consider an $L \times L$ square lattice where
  each node $i$ represents an   agent and the links
represent the connections among agents.  Each agent is connected 
to his four nearest neighbours, the ones above, below,  left and
right. Periodical boundary conditions are adopted so that
the agents at the top  of the lattice are  linked  to those at  the bottom  and
the agents at the left side are linked to those at the right side. 
In this way   the lattice assumes the topology of a torus.

We start with  each agent   initially owning the same amount of capital
consisting of two assets: 
cash $M_{i}(0)$ and $N_{i}(0)$ units of a single  stock.
At each time step $t$ a given trader, $i$,  chooses an action $S_i(t)$
which can take  one of three values: $+1$ if he buys one unit of the
stock, $-1$ if he sells one unit of the stock, or $0$ if he remains  inactive.
The trades undertaken by each player are bounded by
his  resources (agents can be prevented from buying or selling by, respectively,
a shortage of money or stocks)  plus the constraint that he can buy or sell only one
 unit at a time.

We could have alternatively assumed that traders submit an order
which is a normally distributed random variable.
Allowing agents to submit  larger trade quantities  would
create larger fluctuations in the agents
wealth and  would increase the role of budget constrains.  
 As we discuss in the conclusions, analyzing how the distribution of
agents wealth is affected by the trading process  
 would be an interesting extension of this paper. 
Nonetheless the statistical properties of returns  are not likely to be 
influenced by mean-preserving changes in the distribution of individual
trading orders.

In addition to the traders there is a  market maker whose
role  is to clear the orders and adjust prices. The market maker is
also endowed with an initial level of money and stocks and faces the
same constraints on selling short and buying long as the traders but
does not face the unit trade constraint.

The agents' decision  making is driven by  idiosyncratic noise  and the
 influence of  their nearest neighbours. The former is observed only
 once in each trading period but information about the latter can be
 updated  repeatedly within a period. We shall use $t$ to denote a
 trading period and $\tilde{t}$ to denote intra-period time.
At the beginning of each period, each  agent $i$ receives an
 idiosyncratic  signal $\nu_i(t)$ which describes shocks
to his personal preference and is held constant throughout the
 period. The distribution of  $\nu_i$ is  uniform over the interval
 $[-1,1]$.

In addition, within each time period
 he repeatedly exchanges information with  his nearest neighbours.
We use the term
 ``consultation round'' to denote a round of  intra-period communication
 between agents.  
In each consultation round each agent receives a signal  
 $S_j(\tilde{t})$ from his nearest neighbours.
  Note that each $S_j(\tilde{t})$ denotes a temporary
 choice which can get updated from a consultation round to the next.
 Therefore within each consultation round  
each agent receives an aggregate signal  $Y_i(\tilde{t})$:
\be
Y_i(\tilde{t}) =  \sum_{<i,j>} J_{ij} S_j(\tilde{t}) + A  \nu_i(t) 
\ee
 $<i,j>$ denotes that the sum is taken over the set of nearest neighbours of
agent $i$. 
 $J_{ij}$  measures  the influence that   is exercised on agent $i$ by the action
$S_j$ of his  neighbour $j$;   $J_{ij}$ could in principle be asymmetric 
but in this model only  symmetric cases are considered. 

As is well known in  statistical physics,
the behaviour of the system is affected by the 
different   choices  for the $J_{ij}$ in eq.(1).
If $J_{ij} = 0$ then the  traders actions are uncorrelated
with each other.
Taking  all  $J_{ij} = 1$ leads instead to the Ising Model where,
 at a low level of the  idiosyncratic noise, 
the traders  would  reach the same  selling/buying
decisions and generate very large fluctuations in the stock prices. 
Alternatively,   taking
$J_{ij} = 1$ with probability $p$ and $J_{ij} = 0$ with probability $1-p$  
we would be in the framework of  the bond percolation model (Sahimi, 1994)
considered in Cont and Bouchaud (1999) and Stauffer and Sornette (1999)). 
It is well known that  this model is characterized by a percolation
threshold $p_c$ such that when $p<p_c$ the system  decomposes
 into disconnected clusters of agents. A cluster consists of agents who
mutually imitate each other but there is no communication across
clusters. Above $p_c$ a very large cluster appears
which dominates the system. At $p=p_c$ clusters of all sizes form. 
The number of clusters containing $s$ agents decreases with $s$
as a power law. This is the feature underlying the  power law distribution
of returns in the Bouchaud-Cont model. 
Finally choosing the  $J_{ij}$ as  
Gaussian distributed random variables, we would have  an analogy with
spin-glasses (Mezard et al. (1987)). 
In these systems  the structure of the phase space
is extremely  complicated with many possible  stable and
meta-stable states hierarchically organized. 
In our model the main results will be based on assuming $J_{ij} = 1$
with $p=1$ but cases where $p=0$ and $0<p<1$ will also be considered. 

Under frictionless trading each agent would buy at the slightest
positive signal and sell at the slightest negative one. 
We depart from this benchmark by assuming a trade
friction which leads a fraction of the  agents to being inactive in any time period.
This  friction can be interpreted, for example, as a  transaction cost
which is specific to each agent. Alternatively it could be interpreted
as an   imperfect capacity to access information.
Formally we model  this friction as an individual activation 
 threshold which each agent's
signal must exceed to induce  him to trade.
Each agent compares the signal he receives with his individual threshold,
 $\xi_i(t)$, and undertakes the decision: 

\bear
S_i(\tilde{t}) & =  1   \; \; \; & if \;\; \;   Y_i(\tilde{t})  
\geq  \xi_i(t)  \nonumber \\
S_i(\tilde{t}) & = 0    \;\; \;  & if  \; \; \;  -\xi_i(t) 
< Y_i(\tilde{t}) <   \xi_i(t)  \\
S_i(\tilde{t}) & =  -1  \; \; \; & if \; \; \;  Y_i(\tilde{t})  \leq  
- \xi_i(t)  \nonumber
\eear
The $\xi_i(t)$ are chosen initially from a Gaussian distribution, with
 initial variance $\sigma_{\xi}(0)$ and zero  mean. 
Agents'  heterogeneity enters through the distribution of thresholds. 
The homogeneous traders scenario can be recovered in the limit when
$\sigma_{\xi}=0$.
For each agent
$\xi_i(t)$ is held constant throughout a trading period but is 
adjusted over  time proportionally with movements in the stock
 price.

Initially agents whose
 idiosyncratic signal exceeds their individual thresholds make a
 decision to buy or sell and subsequently  influence their
 neighbours' according to eq.(2).
The decision of each trader is updated
sequentially following the rule in eqs.(1) and (2).
 Holding the value of $\nu_i(t)$
and $\xi_i(t)$ fixed,
$Y_i(\tilde{t})$ and  $S_i(\tilde{t})$ are iterated until they
converge to a final value $S_i(t)$ for each trader.  
We checked that convergence   was always reached in our simulations. 

Once the decision process has converged,  traders place their
 orders simultaneously to the market maker and trade takes place at a
 single price.
The market maker  determines the   aggregate demand, $D(t)$,  and supply, $Z(t)$,
  of stocks at time $t$:
\[D(t)=\sum_{i: S_{i}(t)>0}  S_{i}(t)\qquad
Z(t)=-\sum_{i:  S_{i}(t)<0}  S_{i}(t) \]
and the trading volume: $V(t) = Z(t) + D(t)$,  and
adjusts the stock prices according to:
\be
P(t+1) =  P(t)  \left( \frac{D(t)}{Z(t)} \right) ^{\alpha}
\ee
where
\be
\alpha = a \frac{ V(t)}{L^2}
\ee
 $L^2$ is the number of traders 
and represents the maximum number of stocks that can be traded in any
time step. 

The price adjustment rule reflects  an asymmetric reaction 
of the market maker to imbalance orders placed 
in periods of high versus low activity in the market 
and is consistent with the empirically  observed positive correlation
between absolute price returns and trading volume.

The market  volatility can be defined
as  the absolute value of  relative returns:
\be
r(t) = log \frac{P(t)}{P(t-1)}  \;, \;\;\;\; \sigma(t) = |r(t)|
\ee

Price changes lead to an adjustment of next period's thresholds, 
$\xi_i(t+1)$:
\be
\xi_i(t + 1) = \xi_i(t) \frac{P(t)}{P(t-1)} 
\ee

This endogenously generated  dynamics, where  thresholds  follow the
local price trend,
conserves the  symmetry in the probability of   buying versus selling.
Note  that there is a memory effect in the readjustment
mechanism for  thresholds, i.e.  next period's threshold is
proportional to last period's  one  and not to the  initial one.

The  response of thresholds to prices  
induces  a negative correlation between   trading volume and lagged price changes.  
Empirical results on the price-volume
relashionship are not very clear and seem to differ from market to
market (O'Hara (1995)).
In a  theoretical paper  Orosel (1998) introduced
 an overlapping generation model where market
participation covaries positively with share prices.
This  situation could also be considered in our model by reversing the
threshold adjustment mechanism. Indeed the results of our
model are not affected by this choice.

Our choice of the adjustment mechanism
is motivated primarily  by  the fact that if  $\xi_i$ arise from transaction costs, such
as brokerage commissions, these are proportional to stock prices. 
Moreover empirically (Campbell et al. (1993)), 
an asymmetric  leverage effect, i.e. a larger responses of volatility
 to negative as against positive price changes, has
been observed.  Our model accords  with this observation.
Finally a direct interpretation of the  asymmetric change 
of  trading volume to the direction of price change could 
also be attempted: when
prices fall by a large amount  agents are more likely to
 become aware and to react to it than when prices rise
 or stay constant. Casual empiricism suggests that news of a collapse in
stock prices is given  disproportionate prominence in the media.  

\section{Simulations and Results}

The outcomes of the model for different values of the parameters were
simulated numerically. We centered our analysis on the statistical properties
of  the probability distribution  of stock returns and on the   autocorrelation
of  market volatility.

The dimension of the lattice was set at $L=100$. Each agent was initially  given the
same amount of stocks $N_i(0) = 100$ and of cash $M_i(0) = 100  P(0)$,
where $P(0) = 1$. The market maker was given a number of stocks, $N_m$,
which was a  multiple $m=10$ of the number of traders ($L^2$) and
an equivalent  amount of cash.
During the simulations, a record was kept that  each  agent has
 sufficient cash when he buys a stock and at least one stock when he
sells. This was done to ensure that constraints on selling short and
buying long were never violated. It was also checked that the market
always has sufficient inventories to satisfy demand for both money 
and stocks.

The initial value of the thresholds'  variance was set at
$\sigma_{\xi}(0) = 1$.
 The  coefficient $A$ in eq.(1) was fixed at   $A = 0.2$.

For comparison purposes we deviated from the model as described above
in certain instances. These will be discussed as they arise.
In presenting the results of simulations we  first address the issue
of volatility clustering and then the issue of scaling and multi-scaling
in the distribution of returns. 
\vskip .5cm
\centerline{\bf (a)  Volatility Clustering}
\vskip .5cm
We examined the possibilities of volatility clustering both by varying
 the exponent $\alpha$,
which controls the price adjustment speed in eq.(4),
 and the connectivity between agents, as defined by
the value of $J_{ij}$ in eq.(1).

We  first considered the case where 
$\alpha$ is constant and equal to $1$  and the agents act independently,
i.e. $J_{ij} = 1$ with $p=0$ for all $i$ and $j$. 
In this case volatility clustering emerges but, in contradiction with empirical
observations,  is negatively   correlated with 
trading volume. 
This is shown, for example,  
in fig.(1b) where 
the  return sequence and, superimposed, the correspondent trading volume
is plotted  for   $\alpha = 1$ and $p = 0$.

An explanation of the above result is that 
if  agents do not exchange information and are 
not coupled by a common
external signal their aggregate demand and supply
 would be  the sum of random i.i.d. variables and, in the limit of a large
number of traders,  market returns  would be Gaussian
distributed. Nonetheless, if the number of active traders occasionally becomes 
very small following a significant  increase of the price and   
 the  activation  thresholds, a large
imbalance  in aggregate demand  and supply can  still arise as the
idiosyncratic
noise is not averaged away across the active traders. 
As will be explained later, this can lead to periods of high volatility.

Introducing imitation  among the agents by considering  $p > 0$ 
the model  developed  positive volume-volatility correlations. 
Nonetheless, for $p$
below the percolation threshold $p_c$ the expectation is that
distribution of  returns would  not manifest the right 
 power law decay. For this reason  
only cases  where $p$ is close to $p_c$ were considered. 
In these cases, however, the  system  becomes  very unstable
and prices eventually collapse to zero
 as long as $\alpha \sim 1$ or larger. 
On the contrary by choosing $\alpha $ very small
any imbalance in demand and supply was
dampened away  and clustering did not  emerge   even  if a
high  level of communication among the  agents was allowed ($p \sim 1$).

In any case, instead of trying to tune $\alpha$ to a constant value 
which may give the right asymptotic decay of returns, as suggested by 
Zhang (1999), we  choose $\alpha$ as volume dependent.
The effect was that  of stabilizing the system by reducing the
consequences of extreme  demand/supply ratios  
if these were generated
 by the activity of  only a small fraction of traders. 
Choosing $\alpha$ according to eq.(4) and $p=1$
we were  able to generate volatility clustering 
along with both the  observed positive cross-correlation between price volatility and
trading volume as  shown in fig.(2b) and  the empirically observed value
of $\mu$, the exponent of the returns distribution (fig.(5)).

Examples of  price trajectory are 
shown in  fig.(2a) for the above case and in fig.(1a) for the case $\alpha=1$
and $p=0$.
Note that in both  fig.(1a) and fig.(2a) the price trajectories
display no obvious trend. The spectral density of log-prices 
is approximated by $S(f) \sim f^{-\gamma}$ with $\gamma = 1.6$ in the
case of fig.(1a) and,  consistent with
the prediction of a random walk,  $\gamma \sim 2$ in the case of fig.(2a).

In addition to the cases discussed above we have looked at cases where
thresholds either  do not  exist ($\xi_i(t) = 0$ for all $i$ and all
$t$) or they remain constant over time. None of these cases led to
volatility clustering for any of the choices of $p$. Also we considered
cases where thresholds adjust according to eq.(6), $\alpha$ adjust
according to eq.(4) and $p=0$. Again no volatility clustering was observed.

These results suggest that volatility clustering can emerge despite 
no communication among traders, so long as an alternative mechanism exists 
for creating significant imbalances between supply and demand which can
propagate through time. In our paper, the scenario with 
thresholds adjusting according to eq.(6) 
 and a large and constant value of $\alpha$, the speed of
price adjustment, meets this condition.   
With adjusting thresholds,  
an increase in price in one period  reduces
the number of active traders  in the next period and, as argued above, 
 this can lead to  significant imbalances  between
demand and supply.
If, however, $\alpha$ is small or falls with the decrease
 in trading volume, the
propagation mechanism does not work as the 
response of future price is
damped away. On the other hand assuming that $\alpha$ is constant and 
large leads to a significant price change in the following period 
and volatility
clustering can  emerge. In this scenario, periods of
volatility clustering are associated with small overall trading volume, 
leading to a counter-factual correlation between trading volume and
volatility.

Adding communication between traders to a scenario with adjusting 
thresholds makes
the model capable of reconciling large imbalances  between demand and
supply with large overall trading volumes. In this case, which 
is the main case as far as the results of this paper are 
concerned, all three ingredients: imitation, adjusting thresholds and 
variable speed of price adjustment,  play a role in generating
volatility clustering. Imitation 
implies that even if the idiosyncratic signal of an agent is below his  threshold,
 the effect of imitating his neighbours' actions can  
induce his  to trade in the market. Indeed, with imitation, an agent could be induced
to act in a direction opposite to his own idiosyncratic noise signal. 
This imitative behaviour can spread 
through the system from one consultation round to the next,
generating avalanches of different sizes in supply, demand and overall
trading volume.

 While imitation can generate a large spike in trading
volume, adjusting thresholds and a variable speed of price adjustment
help propagate it through time.  Adjusting thresholds create a memory
effect in the overall trading volume, keeping it high or low over several
trading periods. During periods of high activity
large demand/supply imbalances are created through the effects of
imitation and
their impact on future prices is 
 amplified through the rise in the exponent of the price adjustment rule.
 On the other hand, if the trading activity is low, large demand/supply
imbalances are mainly driven by the failure of the law of large
numbers and their effect on future prices is damped away by the fall in
the exponent of the price adjustment rule.
Therefore taking all three components of the market mechanism together
creates volatility clustering along with a positive volatility-volume
correlation,
i.e. bursts of high volatility  arise with  large trading volumes
and periods of low volatility  accompany   low trading volumes.

Volatility clustering indicates the presence of memory effects in the
absolute returns.
A long-memory process  is characterized by 
an hyperbolic decay of its autocorrelation function.
In fig.(3) we plot the autocorrelations function of return $C_{r}$ and  absolute
return $C_{|r|}$  
\bear
C_{r}(L) &= & E(r(t) r(t+L)) - E(r(t))E(r(t+L))   \\
C_{|r|}(L) &= & E(|r(t)| |r(t+L)|) - E(|r(t)|)E(|r(t+L)|) \nonumber    
\eear
as a function of the time lag $L$ for the  case  $p=1$ and $\alpha$ varies
according to eq.(4).
Fig.(3) show that while actual returns are not correlated, the autocorrelation
functions of  absolute returns has a slowly
decaying tail revealing the presence of long term memory. 
$R/S$ analysis provides a 
precise test for inferring whether the decay  of $C_{|r|}(L)$   is
exponential (as in the GARCH model)
or hyperbolic, i.e.
\be
C_{|r|}(L) \sim L^{2H -2}
\ee
where $H$ is the Hurst exponent.
Starting from Mandelbrot (1972), several authors have advocated
 the $R/S$ statistic as  general test of long term memory.
However, Lo (1991)  pointed out that the simple $R/S$ statistic may have
 difficulties in distinguishing between long-memory and short-term
dependence. Given a time series  $X_t$,  Lo (1991) suggested to calculate
 the following modified  $R/S$ (MRS) statistic $Q(s,L)$:
\be
Q(s,L) = \frac{1}{\sigma^*(s,L)}[\; \max_{1 \leq u \leq s}
	\sum_{i = 1}^{u}(X_i - \bar{X}_s)
	 - \min_{1 \leq u \leq s}\sum_{i = 1}^{u}(X_i - \bar{X}_s\;)]
\ee 
where
\be 
\nonumber 
\sigma^*(s,L) = \frac{1}{s} \sum_{i=1}^{s} (X_i - \bar{X}_s)^2 + 
	\frac{2}{s} \sum_{j=1}^{L} \omega_j(L) \sum_{i=j+1}^{s}
	 (X_i - \bar{X}_s) (X_{i-j} - \bar{X}_s)
\ee
$$ = \gamma_o(s) + 2 \sum_{j=1}^{L} \omega_j(L) \gamma(s),  \;\;\; q < n $$

The weights $w_j(L)$ used are 
\be
 \omega_j(L) = 1 - \frac{j}{L+1}
\ee
$\gamma_j$ are the auto-covariance operators calculated up to a lag 
$L = \sqrt{n}$. $\bar{X}_s$ is the mean over a sample of size $s$.
  The case $L = 0$ gives the classical $R/S$ statistic.
 The Hurst exponent, H, is calculated  by a simple linear regression of
$\log(Q(s,L))$ on $\log(s)$. 
If only  short memory  is present $H$ should converge to $1/2$ while with
long memory, $H$ converges to a value larger than
$1/2$.
 We divided our original volatility  sample of  size $n
= 50000$ into $n/s$ non overlapping intervals of size $s$. We
estimated $Q(s,L)$  for each of the  intervals so defined 
and averaged it over all of them. Errors were estimated 
as the standard  deviation of the $Q(s,L)$ over the $n/s$ intervals. 
We repeated this procedure for different 
 values of $s$ in the range $2^4 < s < 2^{11}$.
Results of the MRS statistics  are plotted in fig.(4).
We found  a slope  $H = 0.85$  over the whole  range 
 of considered values of $s$ which
 indicates an hyperbolic decay in the volatility  autocorrelations, with
an exponent $\beta = 2H - 2 =- 0.3$.
 We tested this value  of $\beta$ against 
the plot of the absolute returns autocorrelations
in fig.(3). The solid  line is a power
law curve with exponent $\beta =- 0.3$. The agreement with the data is
good throughout  the  considered values of $L$, up to  $L = 500$. 
Empirical studies (Ding et al. (1993), Cont et al. (1997),
Baviera et al. (1998), Pasquini \& Serva (1999), Liu et al. (1999))
 have estimated a value of
$0.1 < \beta < 0.4$ for the  absolute returns autocorrelation of many
indices and currencies.
Our results are in good quantitative agreement with the empirical
observations.

\vskip .5cm
\centerline{\bf (b) Scaling and Multi-scaling analysis}
\vskip .5cm

In the following we concentrate on the scaling and multi-scaling analysis
for  the simulated data of the main case studied, i.e. 
 when $p = 1$ and $\alpha$ is chosen
accordingly to eq.(4).

We first compare the distribution of returns at different time scales, $\tau$.
In order to do so we define the normalized return
\be
\tilde {r}_\tau(t) =  \frac{1}{v_\tau} \log \frac{P(t)}{P(t - \tau)}
\ee
where $v_\tau = <r^2_\tau(t)> - <r_\tau(t)>^2$ is averaged over the entire
time series of returns.
We define the cumulative probability
 $\Pi(\tilde{r} > r )$ as the probability of finding a normalized 
return larger than $r$.
In fig.(5) we plot, on a log-log scale,  $\Pi(\tilde{r} > r)$  for the rescaled
returns $\tilde {r}_\tau(t)$ 
at $\tau = 1$ and $\tau = 2^{12}$ (positive and
negative returns were merged together by taking  their  absolute values).
If the return distribution decays as a power law $ prob(r) \sim r^{-(1+\mu)}$
than the cumulative distribution $\Pi(\tilde{r} > r)$  
follows
\be
 \Pi(\tilde{r} > r) \sim \frac{1}{r^\mu}.
\ee
$\Pi(\tilde{r} > r)$,
is  well approximated by an inverse cubic-law  (shown in fig.(5) with
a dashed line) over  a certain range of values of $r$. Both the value of
$\mu$ and the range of values of $r$ over which it holds
are consistent with the empirical results.

A way to detect  multi-scaling in  a  time series  
 is through the analysis of the scaling behaviour
 of the moments of the distribution.
Given a process $X_t$, we  define the structure function as 
\be
F_q(\tau) \equiv <|\delta_\tau X|^q> \sim \tau^{\phi_q}
\ee
where  $\delta_{\tau} X = X_{t+\tau} - X_t$.
Independent random walk models always have a unique scaling exponent
 $h$ (the process is said to be self-affine) and $\phi_q = hq$.
For example in the Gaussian case  $h=1/2$. 
Multi-scaling arises if   $\phi_{q}$ is  a non linear function of $q$.
In this case, the process is called multi-affine or multi-fractal.

We analyzed the moments of the absolute returns distribution for our
simulated data and detected
multi-scaling in analogy with empirical findings (Ghashghaie et
al. (1996), Schmitt et al. (1999), Baviera et al. (1998)).
In fig.(6) we plot, for example,  $\log(F_q(\tau))$ versus $\log (\tau)$ for
$q=2,4,6$. The exponent  $\phi_q$  can be estimated by a simple
regression.
 In fig.(7) we plot $\phi_q$ versus $q$ for  $0 < q < 10$.
 We observe two different regions which indicates multi-scaling.
 By using a step-wise  linear
regression, we estimated the slopes as being 
 $0.47$ for $q < 3$   and  $0.14$ for $q > 3$.
These results are in
qualitative agreement with the empirical analysis (Baviera et
al. (1999))  of the $DM/US$
exchange rate quotes. They also found that for $q < 3$ the slope  is
consistent with the random walk hypothesis, according to which 
it is  $ 0.5$, while for $q>3$ 
the  slope of $\phi_q$ falls to a value of $0.256$.

\section{Conclusions}

This paper has outlined a mechanism which can explain  certain
 stylized facts of stock market returns. 
According to our model synchronization effects, which generate large
fluctuations in  returns, can  arise purely  from communication and
imitation among traders,  even in the absence of an  aggregate exogenous
shock. 
On the other hand  the arrival of aggregate news could  lead to synchronized
action  in the absence of communication and also help explain some
stylized facts. In Iori (1999) we discuss the impact of information
arrival in the present model and show that fat tails, volatility clustering and
positive volatility-volume correlations can be generated  by arrival of
sequentially uncorrelated shocks.
Note that previous studies (Copeland \& Friedman (1987), 
 Andersen (1996),  Brock \&  LeBaron (1996))
 on the effect of news on volatility autocorrelation
have relied on a mechanism  of sequential information arrival. 
Hence, communication among traders and aggregate news serve as
complementary channels through which large   fluctuations in
stock prices  and volatility clustering in a  real market may be explained.  

Many interesting questions which have arisen in other studies 
could also be addressed in the  context of our model.  
One of these is how the trading mechanism affects  the  distribution of
wealth among the traders. Under what conditions is the
trading mechanism capable of increasing the average
level of wealth?  
Also, given an initial flat wealth distribution,
how does it change with time as a result of the trading mechanisms? 
Previous  studies  (Levy \& Solomon (1997)) 
suggest  that, in a stationary state, the 
distribution of wealth follows a well defined  power law.

If the model is expanded
to allow for intra-period trading,  agents who have lower thresholds
could trade first  and subsequently influence  their neighbours with higher
thresholds. Hence, a first-mover advantage could arise for  to agents who
have lower thresholds, in that they could benefit from buying (selling)
at lower (higher) prices.
 The intra period trading mechanism  could serve to  explain
the occurrence of the short-term correlations observed in stock returns.

\section*{Acknowledgements}
 I would like to thank D. Farmer, A. Goenka, 
 M. Lavredakis, T. Lux, S. Markose, G. Rodriguez,  E. Scalas,
 D. Sornette,  D. Stauffer, G. Susinno and  
H. Thomas for interesting discussions 
and especially  C. Hiemstra, S. Jafarey and two anonymous referees
for  valuable comments and suggestions.
All responsibility for errors is mine.  

\vskip 1.5cm \noindent
\section*{References}

\vskip .5cm \noindent
Arthur, W. B., J.H. Holland, B. LeBaron, R. Palmer and P. Taylor, (1997),
Asset pricing under Endogenous Expectations in an Artificial Stock
Market, The Economy as an Evolving Complex System II, Addison-Wesley.

\vskip .5cm \noindent
 Andersen T. G.,  (1996), Return Volatility and trading volume: an
information flow interpretation of stochastic volatility, {\em Journal of
Finance}, Vol L1, 1.

\vskip .5cm \noindent
P.Bak, M.Paczuski and M. Shubik,  (1997), Price variation in a Stock Market with
Many Agents, {\it Physica A} {\bf 246}, 430.

\vskip .5cm \noindent
Bannerjee, A., (1992), A simple model of herd behavior,
{\it Quarterly Journal of Economics}, {\bf 107}, 797-818.

\vskip .5cm \noindent
Bannerjee, A., (1993), The economics of rumors,
{\it Review of  Economic Studies}, {\bf 60}, 309-327.

\vskip .5cm \noindent
Baviera R.,
M.~Pasquini, M.~Serva, D.~Vergni, A.~Vulpiani, (1998),
Efficiency in foreign exchange markets,
preprint, http://xxx.lanl.gov//cond-mat/9811066

\vskip .5cm \noindent
 Bollerslev T., (1986),  Generalized Autoregressive Conditional
 Heteroskedasticity, {\it Journal of Econometrics}
{\bf 31}  307-327.
 
\vskip .5cm \noindent
Bouchaud, J.P. and M. Potters, (1999),  Theory of Financial Risks, Cambridge
University Press.

\vskip .5cm \noindent
Brock W. A. and B.D.LeBaron,  (1996), A Dynamical structural model for stock return
volatility and trading volume, {\em Review of Economics and Statistics},
 78, 94-110.

\vskip .5cm \noindent
 Caldarelli G., M.Marsili, Y.C.Zhang, (1997), A prototype model of Stock
Exchange, {\it Europhysics Letters}, {\bf 40}, 479.

\vskip .5cm \noindent
Campbell. J. S. Grossman and J. Wang, (1993), Trading Volume and Serial 
Correlations in Stock Returns, {\em The Quarterly Journal of Economic},
November, 905-939.

\vskip .5cm \noindent
Cont, R. M. Potters, J.P. Bouchaud, (1997), Scaling in stock market
data: stable laws and beyond, 
http://xxxx.lanl.gov//cond-mat/9705087

\vskip .5cm \noindent
Cont, R., J.P.Bouchaud, (1999),
Herd behaviour and aggregate fluctuation in financial markets, 
To appear in:  Macroeconomics dynamics, 

\vskip .5cm \noindent
Copeland T. E. and D.Friedman,  (1987), 
The effect of sequential information arrival on asset prices: an
experimental study.
{\em Journal of Finance}, {\bf XLII}, 3, 763.

\vskip .5cm \noindent
De~Lima P. and N.~Crato, (1994), 
Long range dependence in the conditional variance of stock returns,
{\it Economic Letters} {\bf 45},   281.

\vskip .5cm \noindent
Ding Z., , C.W.J.~Granger, R.F.~Engle, (1993),
A long memory property of stock market returns
and a new model,
{\it Journal of Empirical Finance} {\bf 1}  83.

\vskip .5cm \noindent
Engle, R., (1995),  ARCH: selected readings, Oxford: Oxford University
Press.

\vskip .5cm \noindent
Farmer J. D., (1998),  Market Force, Ecology, and Evolution, 
http://xxx.lanl.gov//adapt-org/9812005.

\vskip .5cm \noindent
Focardi, S. S. Ciccotti and M. Marchesi, (1999), Self-Organized Criticality and
Market Crashes, paper presented at the 4th Workshop on Economics with
Heterogeneous Interacting Agents.

\vskip .5cm \noindent
 Galam S.,  (1997),
Rational group decision making: A random Field Ising Model at $T=0$,
{\em Physica A}, {\bf 238}, 66.

\vskip .5cm \noindent
Ghashghaie S., W. Breymann, J. Peinke, P. Talkner and Y. Dodge, (1996),
Turbulent Cascades in foreign exchange markets, {\em Nature}, 381, 767-770. 

\vskip .5cm \noindent
Gopikrishnan, P.,  V.  Plerou, M.  Meyer, L.A.N.  Amaral  and
H.E. Stanley,  (1999),  Scaling of the distribution of fluctuations of
financial market indices. http://xxx.lanl.gov//cond-mat/9905305.

\vskip .5cm \noindent
Guillaume, D.M.,  Dacorogna M.M., Dav\'e R.R., M\"uller U.A., Olsen
R.B.,  (1997), From the birds eye to the microscope:
 a survey of new stylized facts of the intra-day foreign exchange
markets, 
{\it Finance and Stochastic}, {\bf 1}, 95-130.

\vskip .5cm \noindent
 Hurst,H., (1951), Long term storage capacity of reservoir,
{\em Transactions of the American Society of Civil Engineers} 116, 770-779.
 
\vskip .5cm \noindent
Iori, G. (1999)
 Avalanche Dynamics and Trading Friction effects on Stock Market 
Returns, {\em International Journal of Modern Physics C},
 10, 6, 1149-1162,  1999.

\vskip .5cm \noindent
Levy, M. \& Solomon, S. (1997), 
New Evidence for the Power Law Distribution of Wealth,
{\it Physica A}, {\bf 242}, 90-94. 

\vskip .5cm \noindent
LeBaron B., (1999),  Agent-Based Computational Finance: suggested Reading and
Early Research, to appear in {\em International Journal of Economic Dynamics
and Control}.

\vskip .5cm \noindent
Liu, Y., P. Gopikrishnan, P. Cizeau, M. Meyer, C. Peng, H.E. Stanley
(1999)
The statistical Properties of volatility of price fluctuations, 
http://xxx.lanl.gov//cond-mat/99033609

\vskip .5cm \noindent
Lo A., (1991), Long term memory in stock market prices,
{\em Econometrica} 59, 1279-1313.

\vskip .5cm \noindent
 Lux T. \& Marchesi M., (1999),
"Scaling and criticality in a stochastic multi-agent model of a
financial market", {\em Nature}, 397, 498-500

 \vskip .5cm \noindent
Mandelbrot, B.B. (1963), The variation of Certain Speculative Prices, 
{\em Journal of Business}, {\bf 36}, 394-419.

\vskip .5cm \noindent
Mantegna R., and H.E. Stanley, (1995) Scaling Behaviour in the Dynamics
of an Economic Index {\em Nature}, {\bf 376}, 46-49.

\vskip .5cm \noindent
Mezard M., G.Paris, M.A.Virasoro,(1987), Spin Glass theory and Beyond,
Word Scientific Publishing.

\vskip .5cm \noindent
O'Hara M., (1995), Market Microstructure Theory, Blackwell. 

\vskip .5cm \noindent
Orl\'ean, A. (1995)
Bayesian interactions and collective dynamics of opinion: Herd
behaviour and mimetic contagion,
{\it Journal of Economic Behavior and Organization}, {\bf 28}, 257-74.

\vskip .5cm \noindent
Orosel G. O., (1998),  Participation costs, Trend chasing
and volatility of stock prices, {\em The Review of Financial Studies}, {\bf
11}, 3.

\vskip .5cm \noindent
 Pagan A., The econometrics of financial Markets,  (1996), {\em Journal
of Empirical Finance}, {\bf 3}, 15102.

\vskip .5cm \noindent
 Pasquini M., M. Serva., Clustering of volatility as a multiscale 
phenomenon,
preprint, http://xxx.lanl.gov//cond-mat/9903334 

\vskip .5cm \noindent
Ramsey, J. B., (1997), 
On the existence of macro variables and of macro relationships,
{\it Journal of Economic Behaviour and Organizations}, 30,  275-299.

\vskip .5cm \noindent
Ramsey J. B.  and Z. Zhang, (1997), The analysis of foreign
exchange rates using waveform dictionaries, {\em Journal of Empirical Finance},
{\bf 4},  341-372.

 \vskip .5cm \noindent
Ronalds G. A.,P.E. Rossi, E. Tauchen,  (1992),  Stock prices and volume,
{\em Review of Financial Studies}, {\bf 5}, 199-242.

\vskip .5cm \noindent
Sahimi, M. (1994),  Application of percolation Theory,
Taylor\&Francis.

\vskip .5cm \noindent
Schmitt S., D. Svchertzer and S. Lovejoy, (1999), Multifractal Analysis
of Foreign Exchange Data, {\em Applied Stochastic Models and Data  Analysis},
15, 29-53.

\vskip .5cm \noindent
Sethna  J. P. et al,  (1993),
{\em Physical Review Letters}, {\bf 70}, 3347.

\vskip .5cm \noindent
 Stauffer, P.M.C. de Oliveira and A. Bernardes, (1999),  Monte Carlo Simulation
of Volatility Clustering in a Market Model with Herding, 
{\it Int. J. of Theor. Appl. Finance}, {\bf 2}, 83-94. 

\vskip .5cm \noindent
Stauffer D., D. Sornette, (1999), Self-Organized Percolation 
Model for Stock Market Fluctuations,
http://xxx.lanl.gov/cond-mat/990229.

\vskip .5cm \noindent
Tauchen, G.E., Pitts M., (1983), The price variability-volume
relationship on speculative markets, {\em Econometrica}, {\bf 51},
485-505, 

\vskip .5cm \noindent
Zhang, Y. C., (1999),  Toward a Theory of Marginally Efficient Markets,
 Physica A 269, 30.

\newpage
\begin{figure}
\vskip 2.cm
\centerline{\epsfxsize 8cm\epsfbox{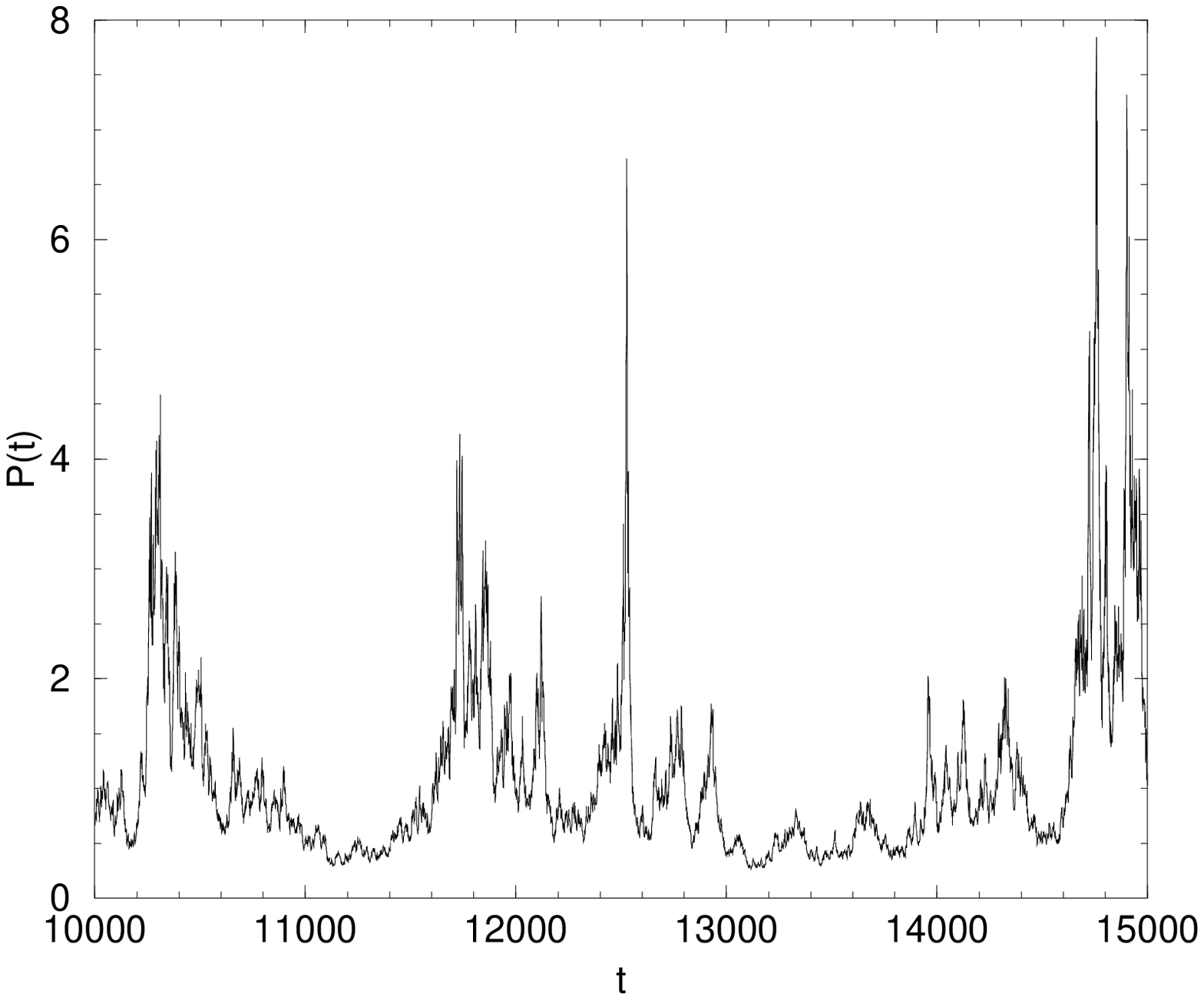}\hskip 1.0cm \epsfxsize 
8cm\epsfbox{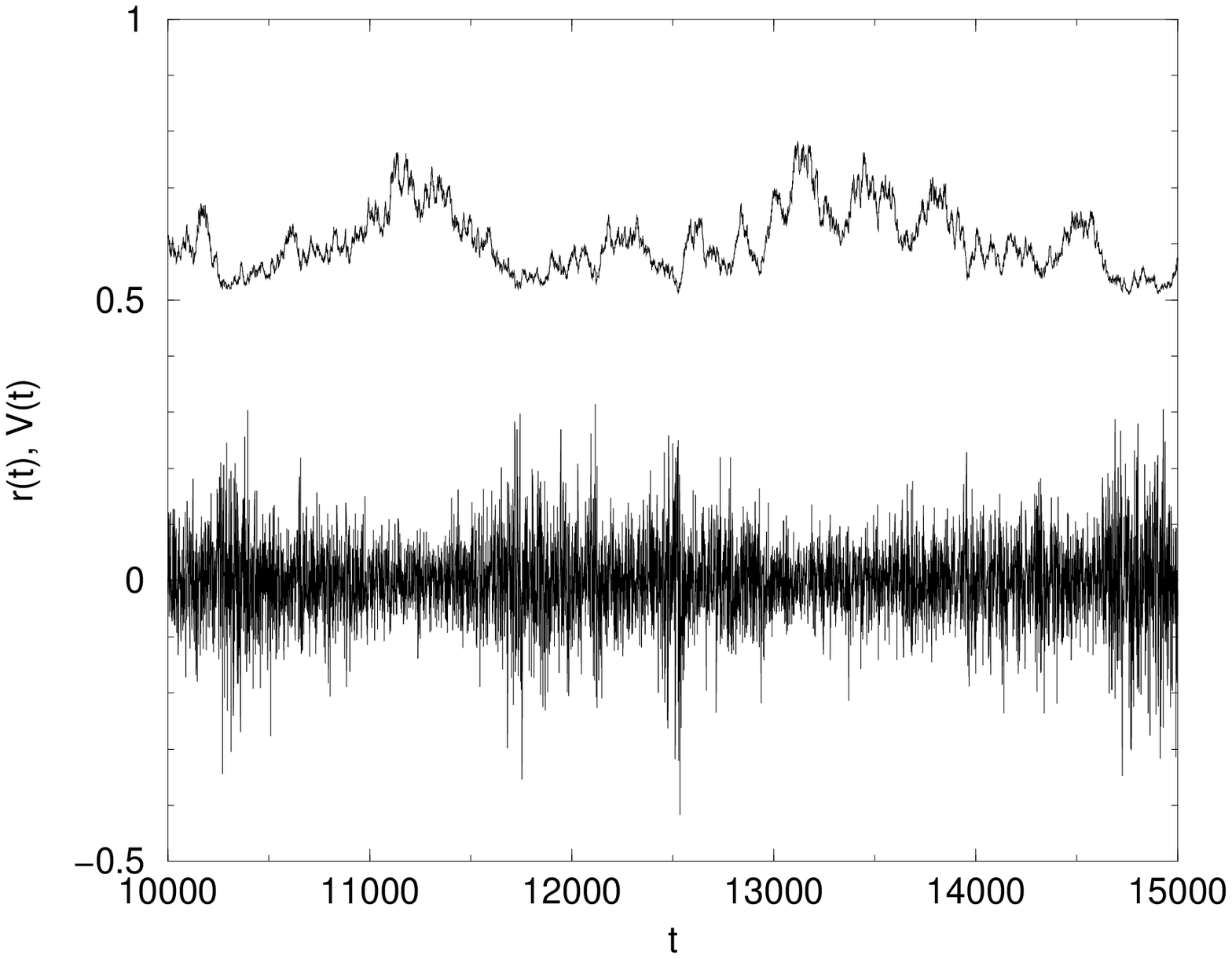}}
\vskip .5cm
  \caption[1]{The two plots correspond to  $\alpha = 1$ and
$p=0$. (a) price history  (b)  return (below) and 
trading volume as a percent of trading agents (above) 
 (the trading volume has been shifted upward by 0.5  for 
a better inspection).} 
  \protect\label{FIG1}
\end{figure}
\newpage
\begin{figure}
 \centerline{\epsfxsize 8.cm\epsfbox{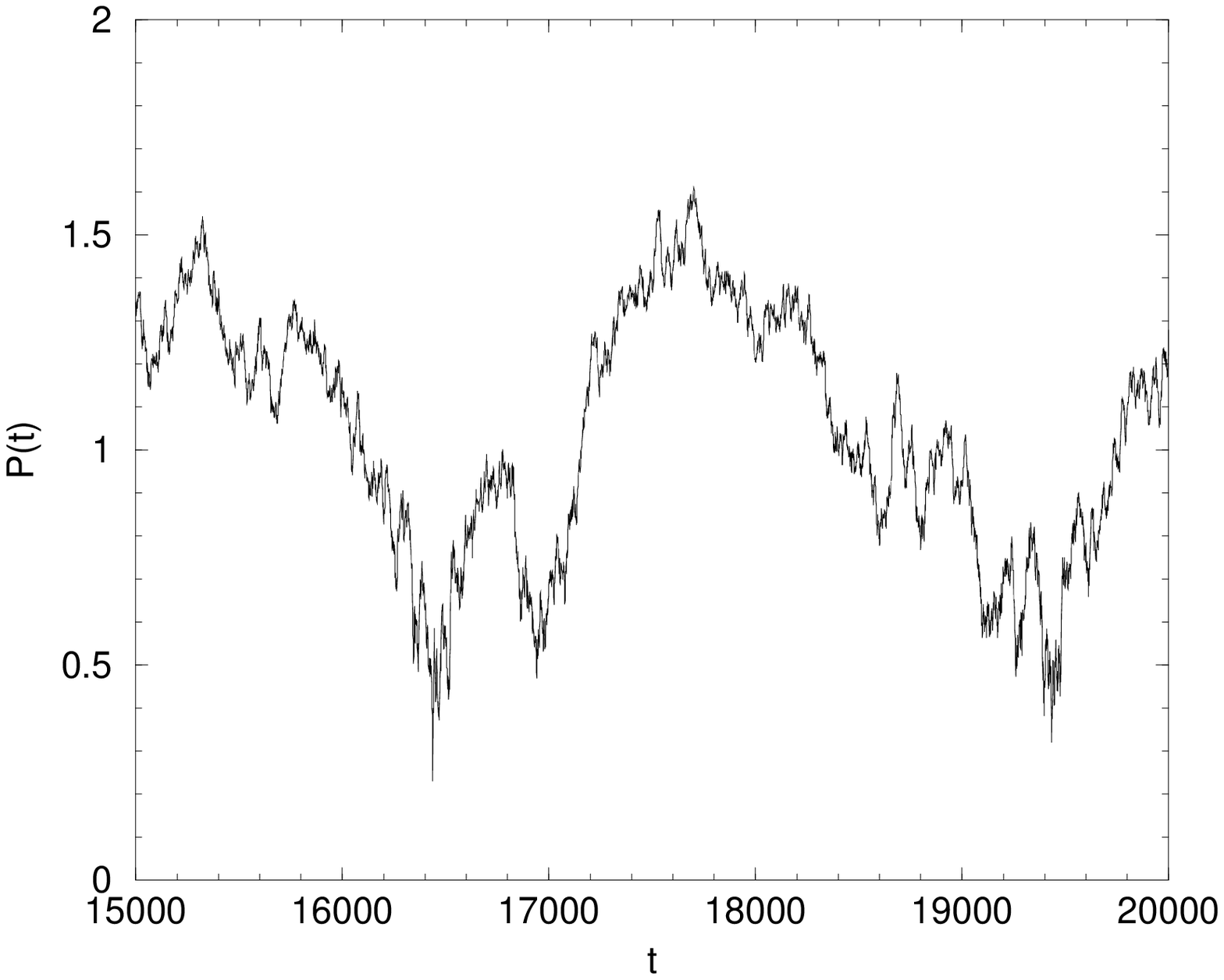}\hskip 1.0cm \epsfxsize 
8.cm\epsfbox{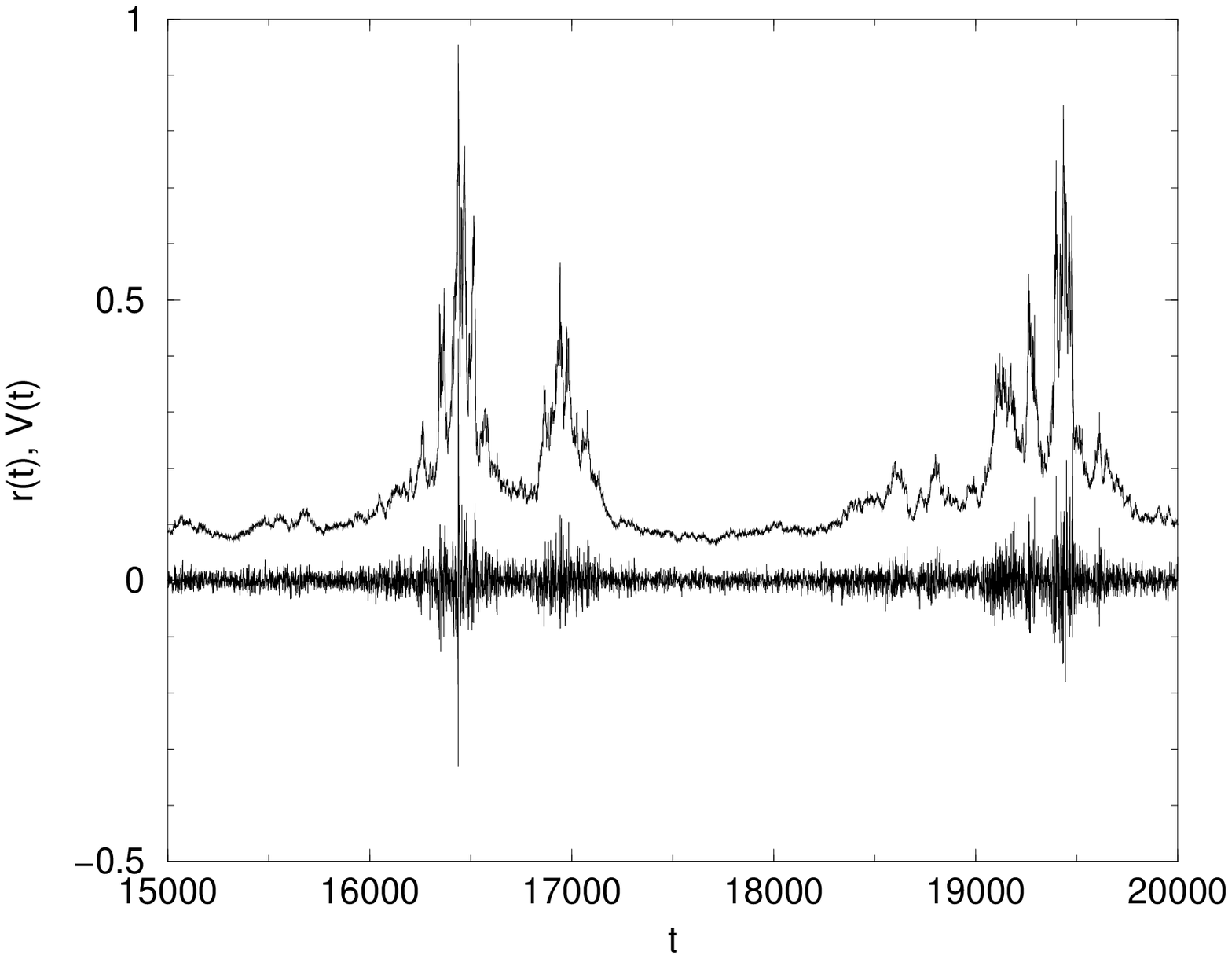}}
\vskip .5cm
  \caption[2]{ The two plots correspond to  $\alpha = a V(t)/N$ and
$p=1$. (a) price history  (b)  return (below) and 
trading volume as a percent of trading agents (above).}
  \protect\label{FIG2}
\end{figure}

\newpage

\begin{figure}
\centerline{\epsfxsize 10cm\epsfbox{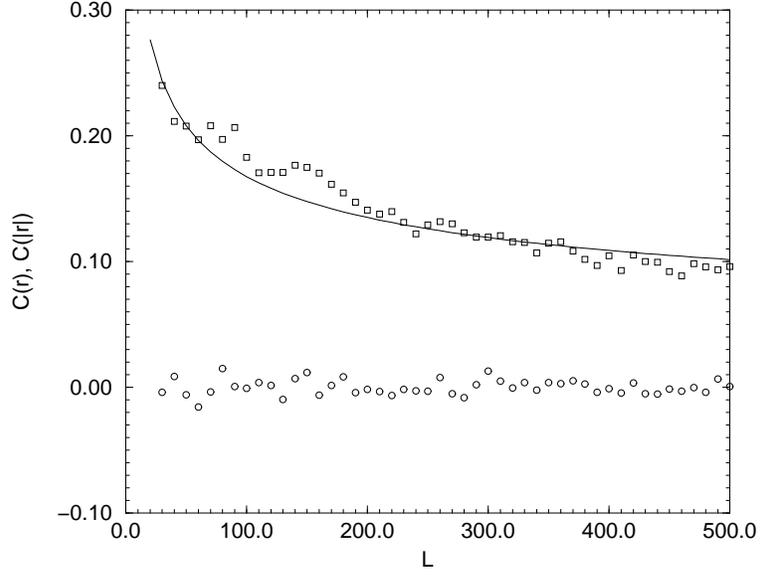}}
\vskip .5cm
  \caption[3]{Autocorrelations function of returns (circles) and  absolute
returns (squares) for  $\alpha = a V(t)/N$ and $p = 1$. The solid line
  is a power law decay with the  exponent 
$\beta=0.3$  estimated from the MRS analysis of absolute returns.}
  \protect\label{FIG3}
\end{figure}

\newpage

\begin{figure}
\centerline{\epsfxsize 10cm\epsfbox{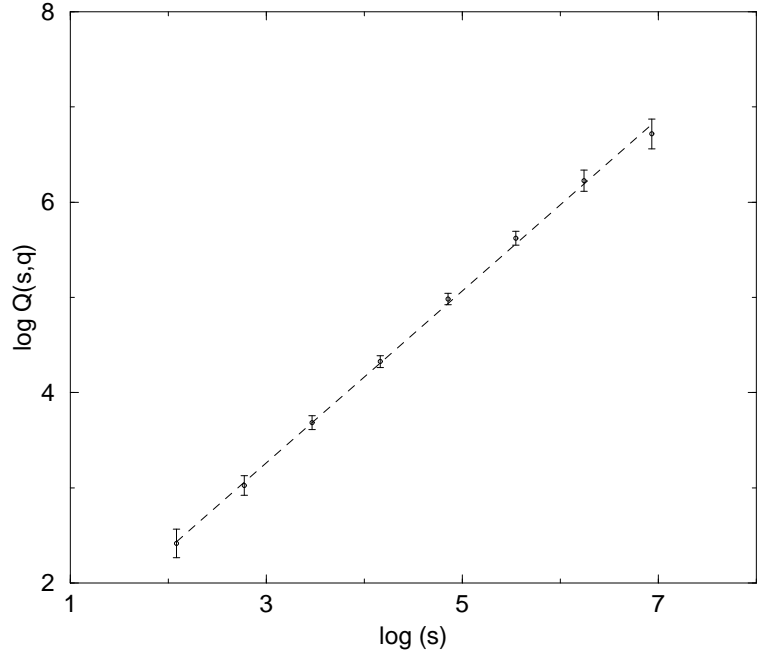}}
\vskip .5cm
  \caption[4]{Modified MRS statistics for absolute  returns
  for
$\alpha = a V(t)/N$ and $p = 1$. We plot the log of 
 $Q(s,L)$, averaged over $n/s$ non overlapping intervals,
 against the log of the sample size $s$.  
Errors  are estimated 
as the standard  deviation of  $Q(s,L)$ over the $n/s$ intervals. 
With a simple linear regression we find
a slope $H = 0.85 $.} 
  \protect\label{FIG4}
\end{figure}
\newpage

\begin{figure}
\centerline{\epsfxsize 10cm\epsfbox{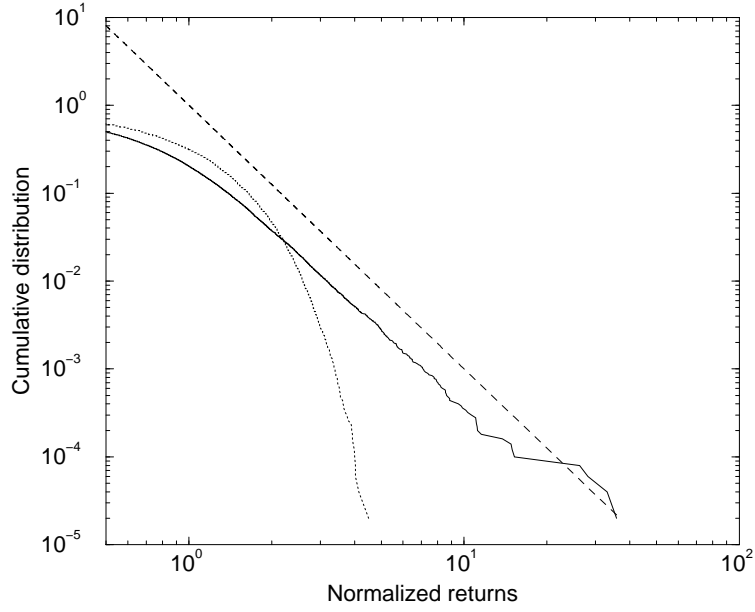}}
\vskip .5cm
\caption[5]{Cumulative distribution of  returns $\Pi(\tilde{r}_{\tau} >
r)$
 at different time lag, $\tau = 1$ (solid), $\tau = 2^{11}$
(dotted). The  dashed  line,  which is for reference, 
 is a power law with an exponent  $\mu = 3$. As can be seen there is a
range of value of $r$ for which the cumulative distribution is parallel to the
reference line indicating consistency with the power law exponent.
 At large $\tau$ the distribution of returns converges to a Gaussian.}
  \protect\label{FIG5}
\end{figure}

\newpage
\begin{figure}
\centerline{\epsfxsize 10cm\epsfbox{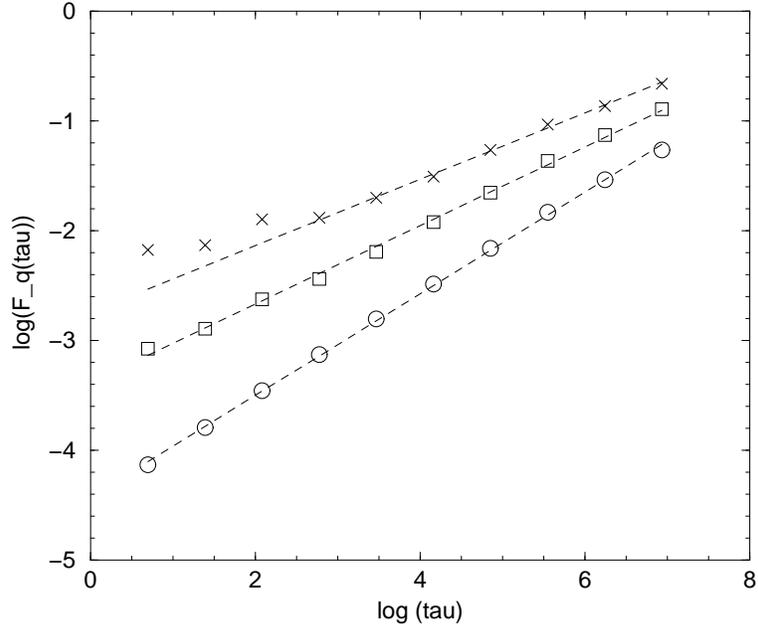}}
\vskip .5cm
  \caption[6]{Structure functions $\log(F_q(\tau))$ versus $\log(\tau)$.
 The three plots correspond to $q=2$ (circles), $q=4$ (squares)
  and $q=6$ (crosses).}
  \protect\label{FIG6}
\end{figure}

\newpage
\begin{figure}
\centerline{\epsfxsize 10cm\epsfbox{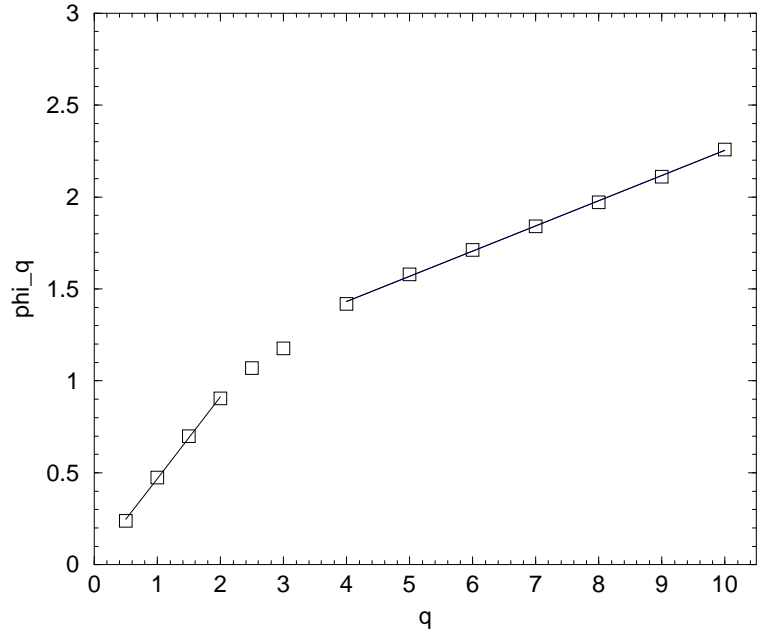}}
\vskip .5cm
  \caption[7]{Exponent $\phi_q$ versus $q$. The slope of the two regions
  are 0.47 at $q < 3$ and  0.14 at $q > 3$.}
  \protect\label{FIG7}
\end{figure}

\end{document}